\documentclass[12pt]{iopart}
\usepackage{graphicx}
\usepackage[dvips]{pstcol}
\usepackage{pst-node}
\usepackage{pst-plot}
\usepackage{pst-coil}
\usepackage{subfigure}
\begin{document}
\title{CP violating asymmetries in the flavour changing single top quark production }
\author{A T Alan, A Senol and A T Tasci}
\address{Department of Physics, Abant Izzet Baysal University, 14280 G\"{o}lk\"{o}y, Bolu, Turkey}
\ead{alan@ibu.edu.tr}
\date{\today}
\begin{abstract}\\CP violating effects in the single top quark production via
flavour changing neutral current (FCNC) reactions
$e^+e^-\rightarrow t\bar{q}$ and $e^+e^-\rightarrow \bar{t}q$
(here $q$ refers to charm and up quarks) are studied. The
effective Lagrangian description of the FCNC interactions is used.
A numerical analysis is performed for some next linear colliders.
CP violating asymmetries for the number of $q$ and $\bar{q}$
quarks are obtained to be of order $10^{-2}-10^{-3}$ depending on
the CM energy.
\end{abstract}

\section{Introduction}
One of the major goals of the next linear colliders (NLC) will be
searching for flavour changing processes. Within the standard
model (SM) these processes occur at the one-loop level and are
unobservably small \cite{1}. Thus any signal of such reactions
will be a clear evidence of new physics beyond the SM. The top
quark with its large mass seems to play the leading role to
provide useful information on new physics. Searching for CP
violation in top sector seems especially promising \cite{top}. For
detailed analyzes of CP asymmetries in single top production
within the minimal supersymmetric model (MSSM) see \cite{2}. On
the flavour changing top-charm (or top-up) transitions, there are
lots of theoretical studies in the literature. Two popular models
allowing such transitions are multi-Higgs-Douplet models \cite{3}
and Supersymmetry \cite{4}. Another possible manifestation of
these kind of new interactions in the top sector is to alter its
couplings to the other known particles, which is called as
effective Lagrangian description \cite{5,6}. In this approach,
deviations from the SM for the flavour changing vertices are
described by a linear effective Lagrangian containing a series of
effective operators whose coefficients are suppressed by power of
$1/\Lambda$, where $\Lambda$ is a high mass scale up to which the
effective theory is assumed to hold. For the flavour changing
processes, the lowest dimension gauge invariant operators built
from SM fields are dimension six, but after spontaneous symmetry
breaking the effective Lagrangian induces the dimension five
operators \cite{5},
\begin{eqnarray}
 {\cal L} & = & ee_q\bar t {i\sigma_{\mu\nu}q^\nu\over
\Lambda}
(\kappa_\gamma-i\tilde\kappa_\gamma\gamma_5)c A^\mu \\
& & +{g\over 2\cos\theta_W}\bar t\left[
\gamma_\mu(v_Z-a_Z\gamma_5)+ {i\sigma_{\mu\nu}q^\nu\over
\Lambda}(\kappa_Z-i\tilde\kappa_Z\gamma_5)\right] cZ^\mu
+h.c.\,,\nonumber \end{eqnarray} where $q$ is the momentum of the
exchanged gauge boson, $\theta_W$ is the Weinberg angle, $e$ and
$g$ denote the gauge couplings relative to $U(1)$ and $SU(2)$
symmetries respectively, $e_q$ denotes the electric charge of
up-type quarks, $A^{\mu}$ and $Z^{\mu}$ the fields of the photon
and $Z$ boson.

 In this paper, we investigate the CP asymmetries in the FCNC
 single top quark production. More precisely, we consider the two
 CP conjugate flavour changing processes $e^+e^-\rightarrow t\bar{q}$ and
$e^+e^-\rightarrow \bar{t}q$ (here q refers to charm or up
quarks). For numerical evaluations, the parameters of the three
high energy linear colliders (LC) LEP II, TESLA and CLIC are used
\cite{7,8}.
\section{Derivation of the CP violating asymmetry}
 Considering only the CP violating $\tilde{\kappa}_{\gamma}$ and
 $\tilde{\kappa}_{Z}$ terms in the effective Lagrangian, contribution of the CP conserving couplings to the asymmetry are less than 0.3 \%,
 the top decays
 $t\rightarrow q\gamma$ and $t\rightarrow q Z$ are written as

\begin{eqnarray}
\Gamma(t\rightarrow q\gamma)&=&\frac{\alpha
e_q^2\tilde{\kappa}_{\gamma}^2}{\Lambda^2m_t^3}\left(m_t^2-m_q^2\right)^3~,
\end{eqnarray}
\begin{eqnarray}
\Gamma(t\rightarrow
qZ)&=&\frac{g_z^2\left(m_t^2-M_Z^2\right)^2}{32\pi
\Lambda^2m_t^3M_Z^2}\\
&\times&\left[\left(m_t^2+2M_Z^2\right)\left[(a_Z^q)^2+(v_Z^q)^2\right]\Lambda^2+\tilde{\kappa}_Z^2M_Z^2\left(2m_t^2+M_Z^2\right)\right]\nonumber~
\end{eqnarray}
where $a_Z^q=\frac{1}{2}$ and
$v_Z^q=\frac{1}{2}-\frac{4}{3}\sin^2\theta_W$.

By using the 95 $\%$ C.L. limits
of $BR(t\rightarrow q\gamma)<0.032$ and $BR(t\rightarrow q
Z)<0.33$ \cite{9}, we obtain that while $\tilde{\kappa}_{\gamma}$ is real, $\tilde{\kappa}_{Z}$ is pure imaginary and the
 restrictions on the couplings are as follows:
\begin{eqnarray}\label{eq3}
|\tilde{\kappa}_{\gamma}|<0.28& ~~\mathrm{and}~~
&|\mathrm{Im}\tilde{\kappa}_{Z}|<0.58~.
\end{eqnarray}
In obtaining these restrictions we have taken $\Lambda=m_t$.

To estimate the possible size of the effect, we consider the
following CP-violating quantity :
\begin{eqnarray}
A_{CP}=\frac{\sigma(\bar{t}q)-\bar{\sigma}(t\bar{q})}{\sigma(\bar{t}q)+\bar{\sigma}(t\bar{q})}
\end{eqnarray}
where $\sigma(\bar{t}q)$ and $\bar{\sigma}(t\bar{q})$ are the
total cross sections for the processes $e^+e^-\rightarrow
\bar{t}q$ and $e^+e^-\rightarrow t\bar{q}$, respectively. The
total cross section for the process  $e^+e^-\rightarrow \bar{t}q$
is obtained as
\begin{eqnarray*}
\sigma=\sigma_{\gamma}+\sigma_Z+\sigma_{int},
\end{eqnarray*}
where $\sigma_{\gamma}$, $\sigma_Z$ are the cross sections for $s$
channel $\gamma$ and $Z$ exchange processes, respectively.
$\sigma_{int}$ corresponds to the interference term of the
amplitude. These are given by

\begin{eqnarray}
\sigma_{\gamma}&=&\frac{2\pi\alpha^{2}e_q^2\tilde{\kappa}_{\gamma}^2}{3\Lambda^2s^3}\left[s^3-3sm_t^4+2m_t^6\right]~,
\end{eqnarray}

\begin{eqnarray}
\sigma_{Z}&=&\frac{1}{16\pi
s^2}\frac{g_z^4\left[(a_Z^e)^2+(v_Z^e)^2\right](m_t^2-s)^2}{24\left[(s-M_Z^2)^2+\Gamma_Z^2M_Z^2\right]}\nonumber\\
&\times&\left[(m_t^2 +2s)\left[(a_Z^q)^2+(v_Z^q)^2\right]
+\frac{(\mathrm{Im}\tilde{\kappa}_{Z})^2s}{\Lambda^2}(2m_t^2+s)\right]~,
\\\nonumber\\
\sigma_{int}&=&\frac{\alpha
e_q\tilde{\kappa}_{\gamma}g_z^2v_Z^e(m_t^2-s)^2}{12\Lambda^2s^2\left[(s-M_Z^2)^2+\Gamma_Z^2M_Z^2\right]}\nonumber\\
&\times&\left[3a_Z^qm_t\Lambda(M_Z^2-s)
+\mathrm{Im}\tilde{\kappa}_{Z}(2m_t^2+s)\Gamma_ZM_Z\right]~.
\end{eqnarray}
where $a_Z^e=-\frac{1}{2}$ and
$v_Z^e=-\frac{1}{2}+2\sin^2\theta_W$. The total cross section for
the CP conjugate process $e^+e^-\rightarrow t\bar{q}$ differs only
in the interference term, that is,
\begin{eqnarray*}
\bar{\sigma}_{\gamma}=\sigma_{\gamma}&~~~~&
\bar{\sigma}_{Z}=\sigma_{Z}
\end{eqnarray*}
and
\begin{eqnarray}
\bar{\sigma}_{int}&=&\frac{\alpha e_q
\tilde{\kappa}_{\gamma}g_z^2v_Z^e(m_t^2-s)^2}{12\Lambda^2s^2\left[(s-M_Z^2)^2+\Gamma_Z^2M_Z^2\right]}\nonumber\\
&\times&\left[3a_Z^qm_t\Lambda(s-M_Z^2)+\mathrm{Im}\tilde{\kappa}_{Z}
(2m_t^2+s)\Gamma_ZM_Z\right]~.
\end{eqnarray}
\section{Numerical results}
\begin{figure}
 \centering
\subfigure[]{\label{fig.1a}
  \includegraphics[width=13.7cm]{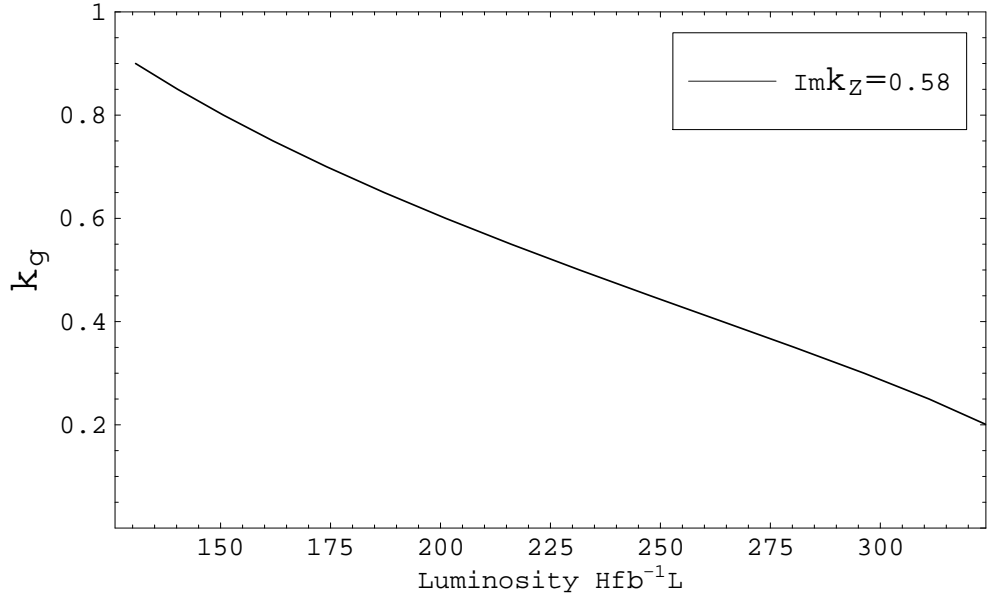}}
\vspace{0.5cm}\subfigure[]{\label{fig.1b}
  \includegraphics[width=14cm]{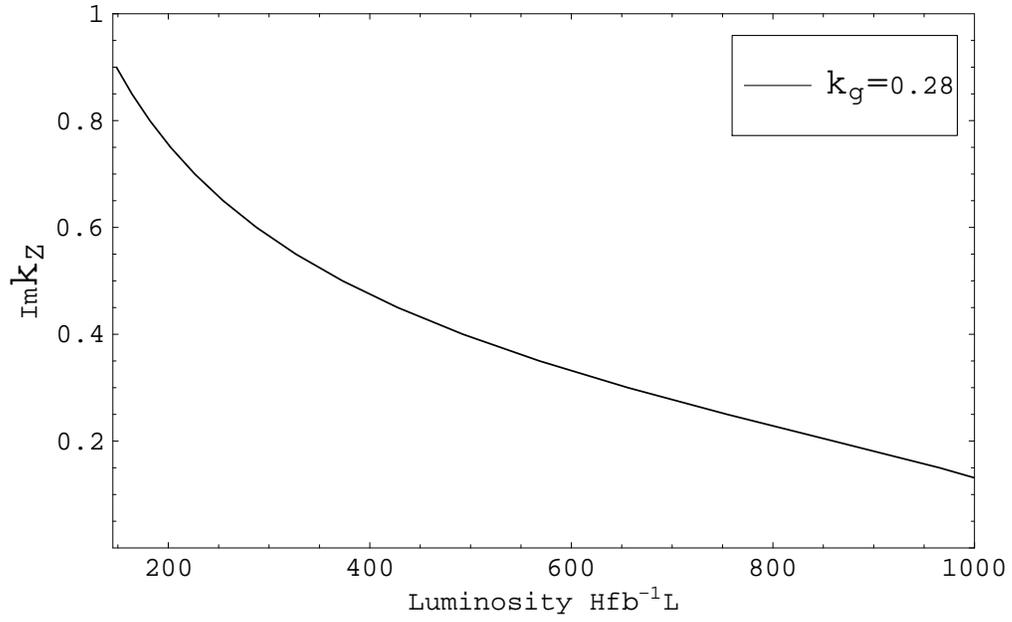}}
 \caption{$95\%$ C.L. sensitivity to the FCNC couplings as a function of the integrated
    luminosity for $\sqrt{s}=500$ GeV~.}
\end{figure}

\begin{figure}[b]
\includegraphics[width=15cm]{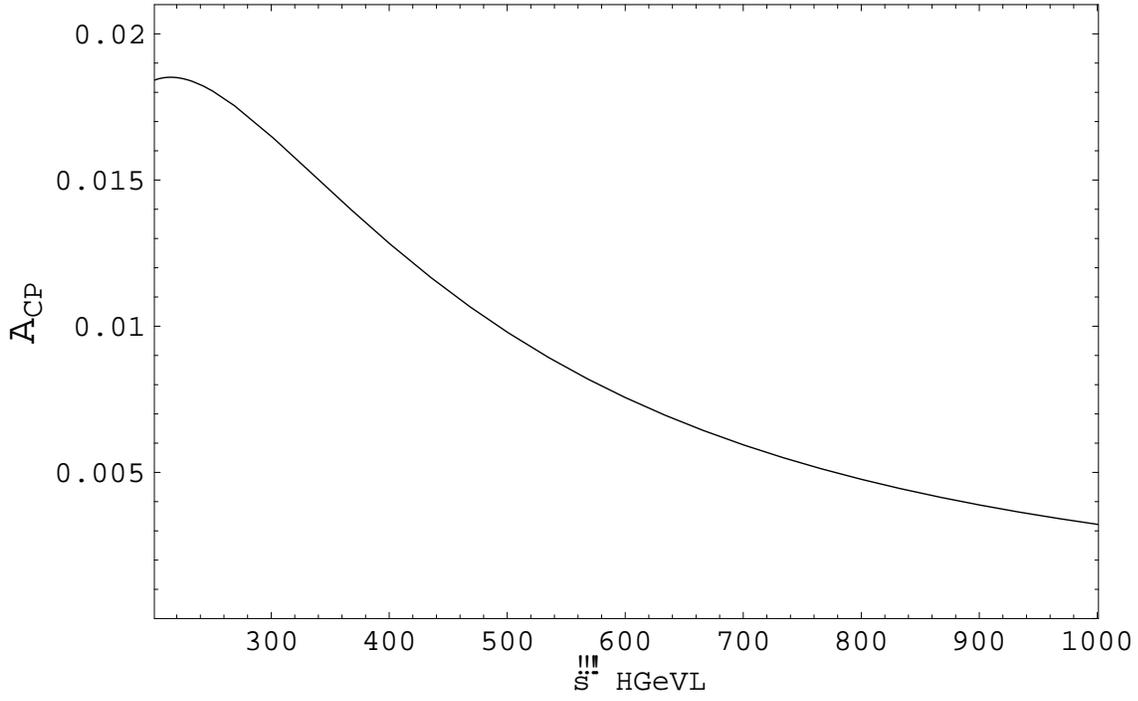}
\caption{\label{fig.2} CP asymmetry as a function of the CM energy
with $\tilde{\kappa}_{\gamma}$=0.28 and
$\mathrm{Im}\tilde{\kappa}_{Z}$=0.58~.}
\end{figure}

\begin{figure}
\centering
  \includegraphics[width=12cm]{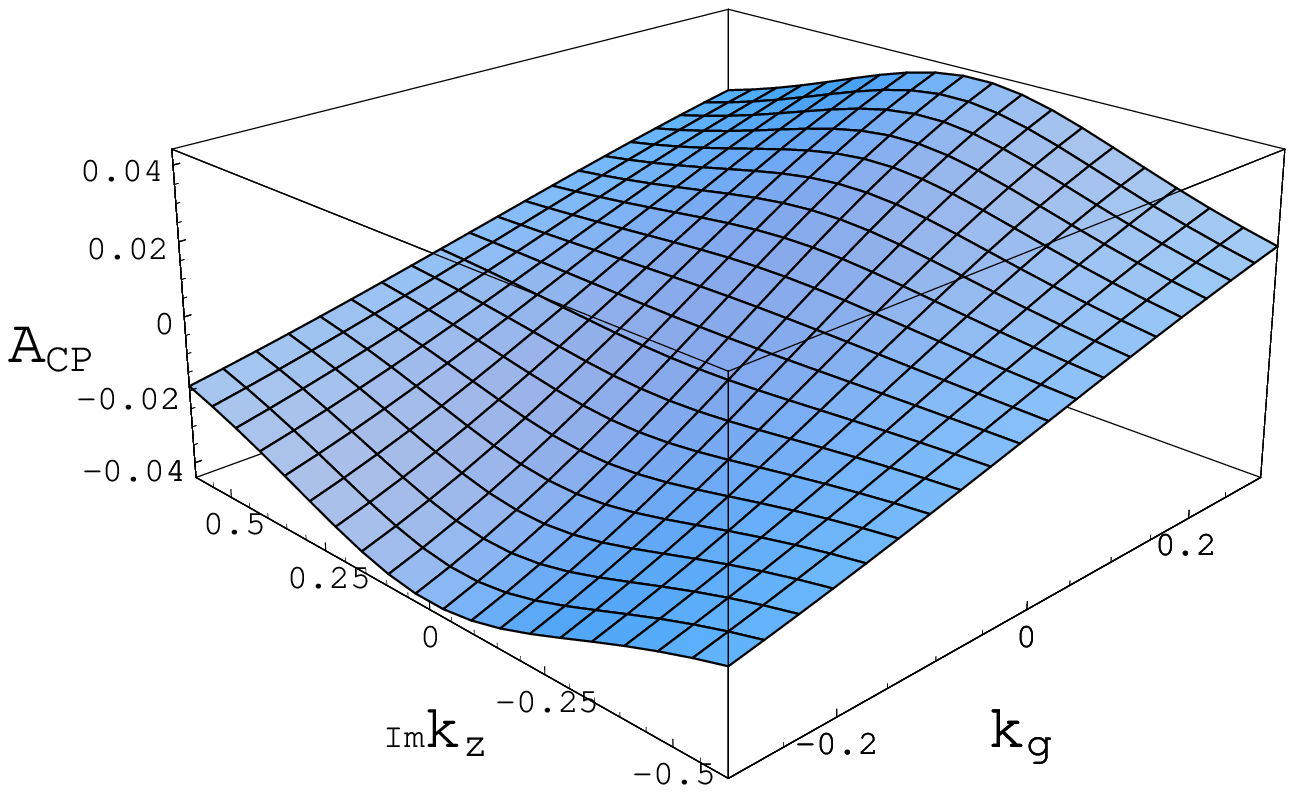}
 \caption{\label{fig.3}CP asymmetry as a function of
        $\tilde{\kappa}_\gamma$ and $\mathrm{Im}\tilde{\kappa}_Z$ at
        $\sqrt{s}=200$ GeV~.}
 \end{figure}

\begin{figure}
\centering
\includegraphics[width=12cm]{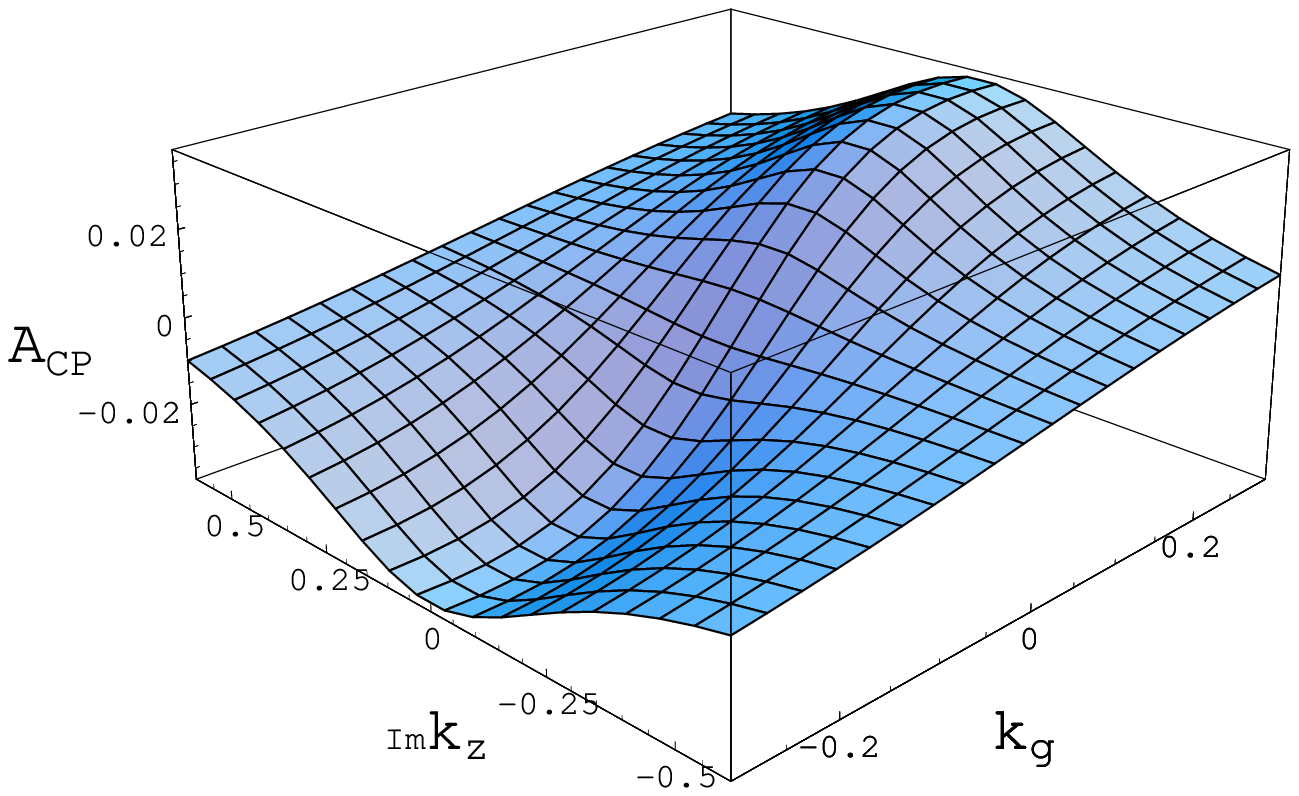}
\caption{\label{fig.4}CP asymmetry as a function of
        $\tilde{\kappa}_\gamma$ and $\mathrm{Im}\tilde{\kappa}_Z$ at
        $\sqrt{s}=500$ GeV~.}
\end{figure}

\begin{figure}
\centering
  \includegraphics[width=12cm]{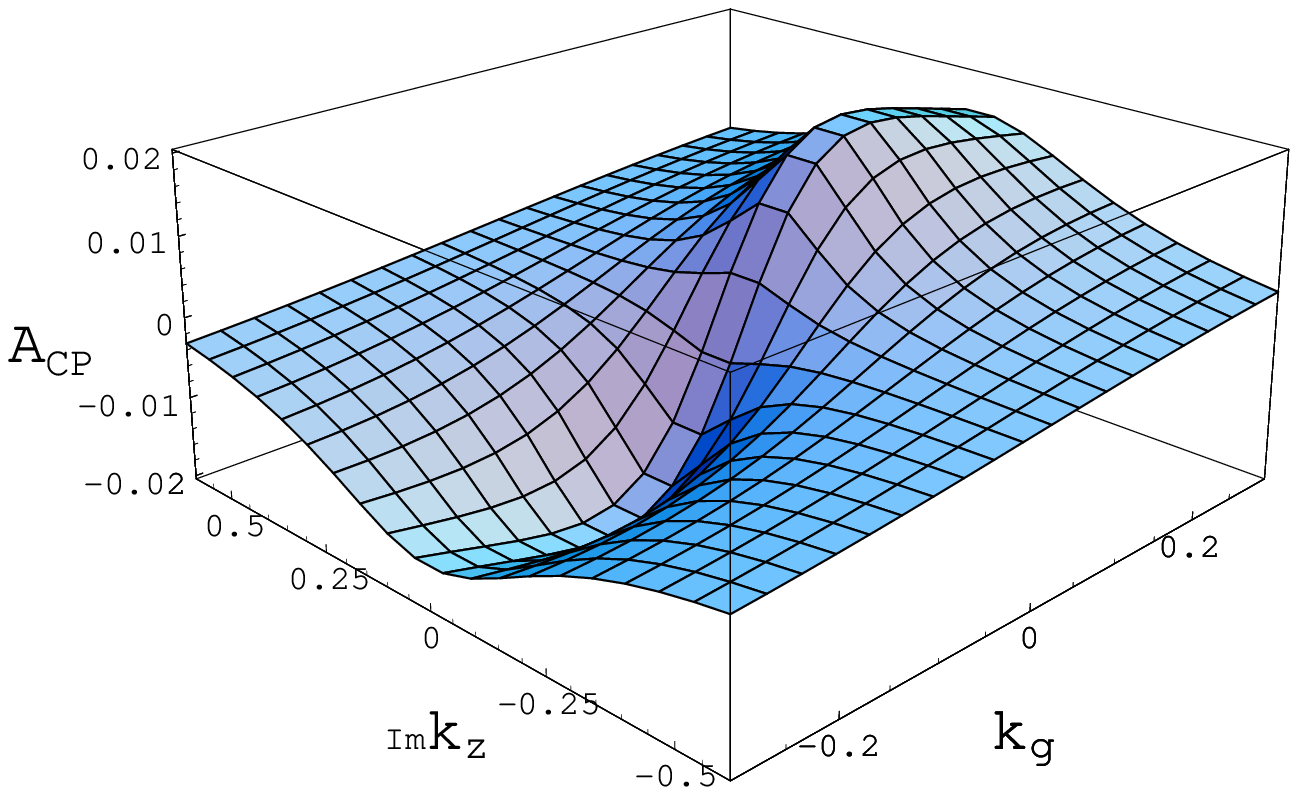}
\caption{\label{fig.5}CP asymmetry as a function of
        $\tilde{\kappa}_\gamma$ and $\mathrm{Im}\tilde{\kappa}_Z$ at
        $\sqrt{s}=1000$ GeV~.}
\end{figure}

The CP asymmetries are evaluated for LEP II with $\sqrt{s}= 200$
GeV, TESLA with $\sqrt{s}= 500$ GeV and CLIC with $\sqrt{s}=$ 1
TeV. With the increasing CM energy, the ability of a collider to
probe the existence of FCNC couplings is improved. But the
sensitivity to these couplings scales with the integrated
luminosity $L$ approximately as $1/\sqrt{L}$. As an example in
figures \ref{fig.1a} and \ref{fig.1b} we display the sensitivity
to the couplings as a function of integrated luminosity at
$\sqrt{s}=500$ GeV. In \fref{fig.2} the asymmetry is displayed as
a function of the CM energy in the range of 200 - 1000 GeV. In
figures \ref{fig.3}, \ref{fig.4} and \ref{fig.5} the CP
asymmetries are shown also as a function of couplings
$\tilde{\kappa}_{\gamma}$ and Im$\tilde{\kappa}_{Z}$ for the CM
energies 200 GeV, 500 GeV and 1000 GeV, respectively. If these
couplings are saturated to their upper bounds given in equation
\eref{eq3}
the following values are obtained for three different $\Lambda$ values: \\

\begin{tabular}{l|lll}
   &$ \Lambda=175$ $\mathrm{GeV}$ & $\Lambda=500$ $\mathrm{GeV}$ &$\Lambda=1000$ $\mathrm{GeV}$
   \\\hline
  $\sqrt{s}=200~\mathrm{GeV}$ & $\mathrm{A}=1.84\times10^{-2}$ & $\mathrm{A}=1.39\times10^{-2}$ & $\mathrm{A}=7.95\times10^{-3}$ \\
  $\sqrt{s}=500~\mathrm{GeV}$ & $\mathrm{A}=9.79\times10^{-3}$ & $\mathrm{A}=1.45\times10^{-2}$ & $\mathrm{A}=1.09\times10^{-2}$ \\
  $\sqrt{s}=1000~\mathrm{GeV}$ &$ \mathrm{A}=3.22\times10^{-3}$ &$ \mathrm{A}=7.17\times10^{-3}$ & $\mathrm{A}=8.16\times10^{-3}$ \\
\end{tabular}
\\\\

\begin{figure}
\centering
  \includegraphics[width=13cm]{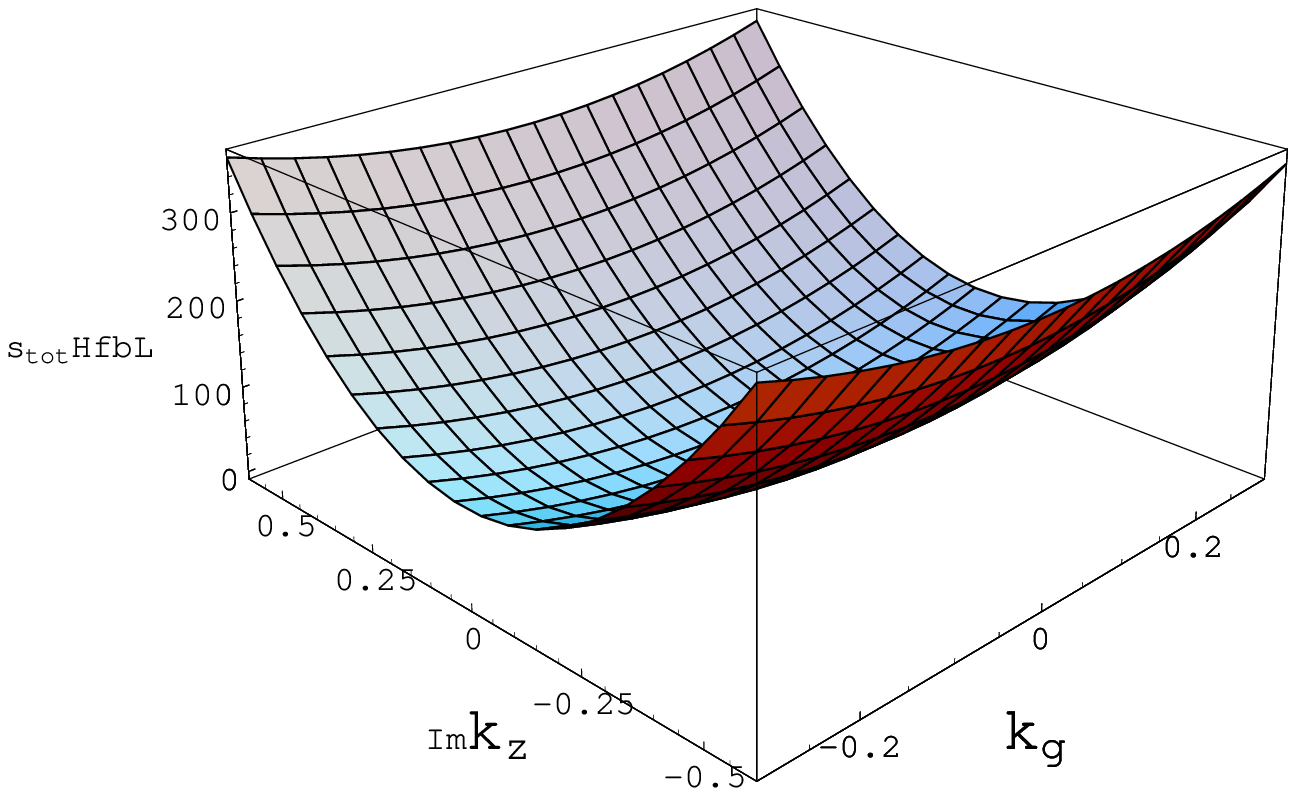}
\caption{\label{fig.6}The total cross-section for
$e^+e^-\rightarrow t\bar{q}$, with $\sqrt{s}=500$ GeV, as a
function $\tilde{\kappa}_\gamma$ and
$\mathrm{Im}\tilde{\kappa}_Z$.}
\end{figure}

The total cross section for the process $e^+e^-\rightarrow
t\bar{q}$, with $\sqrt{s}=500$ GeV, as a function of the couplings
is plotted in figure \ref{fig.6}. In all figures we have taken
$\Lambda=m_t$.
\section{Conclusions}
We have studied the CP asymmetries in single top production in
$e^+e^-$ collisions via FCNC couplings. We used a model
independent effective Lagrangian approach to describe these
couplings. In this description, the effects turn out to be of the
order of $10^{-2}$-$10^{-3}$. The total cross section for
$e^+e^-\rightarrow t\bar{q}$ is $\sim$ 330 fb which means $\sim$
$10^5$ events for TESLA with an integrated luminosity of 300
fb$^{-1}$. Therefore, an asymmetry of the order of $10^{-2}$ is to
be detectable at TESLA. Finally, similar discussions on the
observability of the asymmetries can be done for other future
linear colliders, e.g. for CLIC.
 \ack{ This work was supported
in part by Abant Izzet Baysal University Research Fund.}

\newpage


\section*{References}


\begin{thebibliography}{99}

\bibitem{1}

Ganapathi V \etal 1983 \textit{Phys. Rev.} \textbf{D27} 579

Buchm\"{u}ller W and Gronau M 1989 \textit{Phys. Lett.}
\textbf{B220} 641

Fritzsch H 1989 \textit{Phys. Lett.} \textbf{B224} 423

Huang C S, Wu X H and Zhu S H 1999 \textit{Phys. Lett.}
\textbf{B452} 143

\bibitem{top}
Atwood D, Bar-Shalom S, Eilam G and Soni A 2001 \textit{Phys.
Rept.} \textbf{347} 1

\bibitem{2}

Bar-Shalom S, Atwood D and Soni A 1998 \textit{ Phys. Rev.}
\textbf{D57} 1495

Christova E 1999 \textit{Int. J. Mod. Phys.} \textbf{A14} 1

\bibitem{3}Atwood D, Reina L and Soni A 1997 \textit{Phys. Rev.} \textbf{D55} 3156

Bar-Shalom S, Eilam G, Soni A and Wudka J 1997 \textit{Phys. Rev.
Lett.} \textbf{79} 1217

\textit{ibid} 1998 \textit{Phys. Rev.} \textbf{D57} 2957

\bibitem{4}Li C S, Oakes R J and Yang J M 1994 \textit{Phys. Rev.} \textbf{D49} 293

Atwood D, Reina L and Soni A 1995 \textit{Phys. Rev. Lett.}
\textbf{75} 3800

Bar-Shalom S, Atwood D, Eilam G, Mendel R R and Soni A 1996
\textit{Phys. Rev.} \textbf{D53} 1162

Hou W S and Lin G L 1996 \textit{Phys. Lett.} \textbf{B379} 261

Mahanta U and Ghosal A 1998 \textit{Phys. Rev.} \textbf{D57} 1735

Bar-Shalom S, Eilam G and Soni A 1999 \textit{Phys. Rev.}
\textbf{D59} 055012

Yu Z H \etal 2000 \textit{ Eur. Phys. J.} \textbf{C16} 541

Chemtob M and Moreau G 1999 \textit{Phys. Rev.} \textbf{D59}
116012

\bibitem{5}
Han T and Hewett J L 19999 \textit{Phys. Rev.} \textbf{D60} 074015

\bibitem{6}


Peccei R D and Zhang X 1990 \textit{Nucl. Phys.} \textbf{B337} 269

Han T, Peccei R D and Zhang X 1995 \textit{Nucl. Phys. }
\textbf{B434} 527

Han T, Whisnant K, Young B L and Zhang X 1997 \textit{Phys. Rev}
\textbf{D55} 7241

Bar-Shalom S and Wudka J 1999 \textit{ Phys. Rev.} \textbf{D60}
094016

Tait T and Yuan C P 2001 \textit{ Phys. Rev.} \textbf{D63} 014018

Alan A T and Senol A 2002 \textit{Europhys.Lett.} \textbf{59}
669-673


\bibitem{7}
Assmann R W \etal 2000 \textit{The CLIC Study Team} CERN 2000-008

\bibitem{8} \textit{TESLA Technical Design Report} (2001) DESY
2001-011

\bibitem{9}
Groom D E \etal 2000 \textit{Eur. Phys. J.} \textbf{C15} 1


\end{thebibliography}
\end{document}